\documentclass[journal]{IEEEtran}

\usepackage{cite}

\ifCLASSINFOpdf
   \usepackage[pdftex]{graphicx}

\else

\fi

\usepackage{amsmath}
\usepackage{multirow}
\usepackage[normalem]{ulem}
\newcommand\norm[1]{\left\lVert#1\right\rVert}
\usepackage{float}
\usepackage{tikz,pgfplots,rotating,amsmath}
\usetikzlibrary{colorbrewer}
\usetikzlibrary{patterns}
\usepgfplotslibrary{fillbetween}
\usepackage{pgfplots}
\usetikzlibrary{pgfplots.groupplots}

\usepackage{xcolor}
\usepackage{amssymb}
\usepackage{mwe}
\usepackage{graphbox}
\usepackage{wrapfig}

\newcommand{\correctiondelete}[1]{{\color{blue} \sout{#1}}} 

\begin{document}

\IEEEspecialpapernotice{(Invited Paper)}
\title{Dual-polarisation Non-linear Frequency-division Multiplexed Transmission with $b$-Modulation}

\author{Xianhe~Yangzhang,
        Vahid~Aref,
        Son Thai~Le, Henning Buelow, Domani\c c Lavery and Polina~Bayvel,~\IEEEmembership{Fellow,~IEEE}
\thanks{Manuscript received October 23rd, 2018; revised month day, 2018.}
\thanks{X. Yangzhang, D. Lavery and P. Bayvel are with the Department
of Electrical and Electronic Engineering, University College London, WC1E 7JE London, UK (e-mail: x.yangzhang@ucl.ac.uk). The work of X. Yangzhang is within the COIN project, which has received funding from the European Union's Horizon 2020 research and innovation programme under the Marie-Skłodowska-Curie grant agreement No.676448.}
\thanks{V. Aref, S. T. Le and H. Buelow are with Nokia Bell Labs, 70435 Stuttgart, Germany. The work of V. Aref and S. T. Le has been performed in the
framework  of  the  CELTIC  EUREKA  project  SENDATE-TANDEM  (Project
ID  C2015/3-2),  and  it  is  partly  funded  by  the  German  BMBF  (Project  ID
16KIS0450K).}
}

\markboth{Journal of Lightwave Technology}%
{Yangzhang \MakeLowercase{\textit{et al.}}: Dual-polarisation Non-linear Frequency-division Multiplexed Transmission with b-Modulation}
%



\maketitle

\begin{abstract}
There has been much interest in the non-linear frequency-division multiplexing (NFDM) transmission scheme in the optical fibre communication system. Up to date, most of the demonstrated NFDM schemes have employed only single polarisation for data transmission. Employing both polarisations can potentially double the data rate of NFDM systems. We investigate in simulation a dual-polarisation NFDM transmission with data modulation on the $b$-coefficient. First, a transformation that facilitates the dual-polarisation $b$-modulation was built upon an existing transformation in \cite{yousefi2016ecoc}. Second, the $q_c$- and $b$-modulation for dual-polarisation were compared in terms of Q-factor, spectral efficiency (SE), and correlation of sub-carriers. The correlation is quantified via information theoretic metrics, joint and individual entropy.  The polarisation-multiplexed $b$-modulation system shows 1 dB Q-factor improvement over $q_c$-modulation system due to a weaker correlation of sub-carriers and less effective noise. Finally, the $b$-modulation system was optimised for high data rate, achieving a record net data rate of 400 Gbps (SE of 7.2 bit/s/Hz) over $12\times 80$ km of standard single-mode fibre (SSMF) with erbium-doped fibre amplifiers (EDFAs). Based on the above simulation results, we further point out the drawbacks of our current system and quantify the error introduced by the transceiver algorithms and non-integrability of the channel. 
\end{abstract}
 

%
\IEEEpeerreviewmaketitle

\section{Introduction}
%
%
%
%
\IEEEPARstart{O}{n}e factor limiting the data rate in optical fibre networks is the signal-signal interference induced by Kerr non-linearity \cite{Essiambre2010}. To further improve the data throughput in optical fibre networks, NFDM was proposed to constructively include the non-linear effect into the signal design, undertaking the challenge of achieving higher data rates.   

In an NFDM system, information is encoded on non-linear frequencies, which is defined by the non-linear Fourier transform (NFT) of a time-domain signal. The non-linear frequencies consist of real-valued frequencies and complex frequencies (called eigenvalues). They have the advantage of being independent components in the integrable non-linear Schr\"odinger equation (NLSE) for single polarisation (SP) or Manakov equation for dual polarisation (DP). Interference-free or weak-interference transmission seems possible in a network environment as conjectured in \cite{Yousefi2014}. NFDM suffers significantly less from inter-channel interference than the conventional linear multiplexing schemes as shown in \cite{yangzhang2018jlt,yousefi2016ecoc} utilising only the continuous spectrum which contains real-valued frequencies. The two papers have a transmission scheme of  one symbol per channel-use in a multi-user environment over fibre links either with ideal distributed amplification \cite{yousefi2016ecoc} or with lumped amplification \cite{yangzhang2018jlt}. 

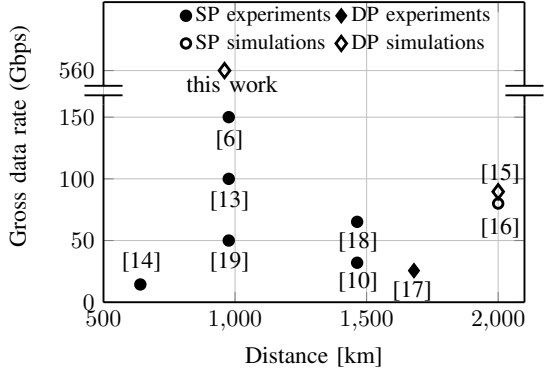
\begin{figure}[t]
\centering
\pgfplotsset{
    every non boxed x axis/.style={} 
}
\begin{tikzpicture}[scale = 0.9]
\begin{groupplot}[    group style={
        group name=my fancy plots,
        group size=1 by 2,
        xticklabels at=edge bottom,
        vertical sep=0pt
    },
xmin=500,xmax=2100,  
width=7.8cm,
height = 4cm,
yticklabel style = {font=\normalsize},
xticklabel style = {font=\normalsize},
xtick={500, 1000, 1500, 2000},
xlabel style={font=\normalsize},
ylabel style={font=\normalsize},
xlabel={Distance [km]},
ylabel style={xshift=1cm},
ylabel={Gross data rate (Gbps)},
axis x line=bottom,
grid,
]
\nextgroupplot[ymin=540,ymax=600,line width=0.8, 
               ytick={560},
              axis x line=top, 
              xlabel = \empty,
               axis y discontinuity=parallel,
               ylabel =\empty,
               height=3.1cm]
\node[draw=none] at (1.9cm, 0.3cm)   () {\textcolor{black}{this work}};
\addplot[ dashed, only marks, name path=lowerbound, line width=1.2, color=black, mark=diamond, mark size=3pt, mark options = {solid}] 
 table[row sep=crcr]{%
960 560\\
 };               
\nextgroupplot
[ymin=0,ymax=160,line width=0.8,
legend entries={\textcolor{black}{\small SP experiments},\textcolor{black}{\small DP experiments},\textcolor{black}{\small SP simulations}, \textcolor{black}{\small DP simulations}},
legend style={at={(0.95,1.55)}, draw=none, legend columns=2, legend cell align=left, fill=none},
height=4.5cm
]
\node[draw=none] at (3.7cm, 0.9cm)   () {\cite{Le:17ofc}};
\node[draw=none] at (1.8cm, 2.4cm)   () {\cite{le2018jlt}};
\node[draw=none] at (1.8cm, 1.5cm)   () {\cite{Le:18}};
\node[draw=none] at (3.7cm, 0.3cm)   () {\cite{Lena2017}};
\node[draw=none] at (1.8cm, 0.6cm)   () {\cite{Le:17jlt}};
\node[draw=none] at (0.5cm, 0.6cm)   () {\cite{Gui:18}};
\addplot[ dashed, only marks, name path=lowerbound, line width=1.2, color=black, mark=*, mark size=2pt, mark options = {solid}] 
 table[row sep=crcr]{%
1464 32\\
1464 65.16\\
976 150\\
976 100\\
976 50\\
640 14.4\\
 };

\node[draw=none] at (4.5cm, 0.17cm)   (a) {\cite{Gemechu2018}}; 
\addplot[ dashed, only marks, name path=lowerbound, line width=1.2, color=black, mark=diamond*, mark size=2.5pt, mark options = {solid}] 
 table[row sep=crcr]{%
1680 25.6\\
 };

\node[draw=none] at (5.8cm, 1.1cm)   (a) {\cite{Civelli:18}}; 
\addplot[ dashed, only marks, name path=lowerbound, line width=1.2, color=black, mark=o, mark size=2pt, mark options = {solid}] 
 table[row sep=crcr]{%
2000 80\\
 };
 
\node[draw=none] at (5.8cm, 1.9cm)   (a) {\cite{Goossens17}}; 
\addplot[ dashed, only marks, name path=lowerbound, line width=1.2, color=black, mark=diamond, mark size=3pt, mark options = {solid}] 
 table[row sep=crcr]{%
2000 89.6\\
 };

\end{groupplot}
\end{tikzpicture}
\caption{Gross data rate of recently implemented NFDM systems of different transmission distances in experiments and simulations.
} \label{fig:recent}
\end{figure}

In the last few years, there has been very active research on the single-channel NFDM transmission systems, such as a few recent transmissions utilising continuous spectrum \cite{Aref2018,le2018jlt,Gemechu2017ecoc,Civelli:18cleo,tavakkolnia2017signaling, Lena2017}. SE up to 2.3 bit/s/Hz \cite{le2018jlt} has been achieved in the SP-NFDM system up to date. For a thorough and up-to-date review, the readers are referred to \cite{Turitsyn:17}.  More recently, it was proposed to modulate information on $b$-coefficients, instead of $q_c$-coefficients, in order to generate time-limited SP-NFDM signals \cite{wahls2017ecoc}. Subsequently, the experiment reported in \cite{Le:18,Gui:18} showed the benefit of $b$-modulation in an SP-NFDM system. The improvement was partially attributed to the compactness of time-domain signals generated by $b$-modulation, i.e., less tail-truncation. 
Employing dual polarisation can potentially double the data rate. A few recent studies on DP-NDFM systems were carried out in \cite{Goossens17, Civelli:18,Gemechu2018} all modulating $q_c$-coefficients. Some recent results are summarised in Fig.~\ref{fig:recent} in terms of gross data rate.

Progress in this work is threefold, i) $b$-modulation was modified to dual polarisation by a transformation built upon an existing one for single polarisation in \cite{yousefi2016ecoc}, ii) comparisons were carried out to show that in a DP-NFDM system, Q-factor in $b$-modulated system is 1 dB higher than in $q_c$-modulated system. This is similarly observed in \cite{Le:18,Gui:18} for single polarisation. To have a deeper insight about the underlying reason, we studied the correlations of information sub-carriers in $q_c$-modulation and in $b$-modulation via the relevant information theoretic metrics of joint and individual entropies. A clear link has been observed between the Q-factor improvement and the weaker correlation of sub-carriers. iii) the $b$-modulated DP-NFDM system was systematically optimised for high data rate. As a result, a record net data rate of 400 Gbps with a SE of 7.2 bits/s/Hz was achieved in simulation over 12 spans of 80 km SSMF with EDFAs. 

The paper is organised as follows: Sec.~\ref{sec:NFT} gives a brief review of the NFT framework in dual-polarisation. Sec.~\ref{sec:qcb} 
describes the $q_c$- and $b$-modulated DP-NFDM systems. The optimisation of $b$-modulated system is carried out in Sec.~\ref{sec:drawbacks}. Sec.~\ref{sec:cov} analyses the correlation of sub-carriers in both $q_c$- and $b$-modulation. Some remarks on the drawbacks of our current system and their quantification are given in Sec. \ref{sec:drawbacks}. Sec. \ref{sec:conclusion} concludes the paper.

\subsubsection*{Notations}
We denote continuous DP-signals as $\vec{Q}(t, z)=[Q_1(t,z)~Q_2(t,z)]$. We denote the Euclidean norm by $\left\Vert\cdot\right\Vert$. A discrete complex random vector of length $N$ and its elements are both written in upper case letter, without and with subscripts, respectively, e.g., $\vec{X}=[X_1~X_2~...~X_N]^T$. A realisation of the complex random vector is denoted in lower case letter such as $\vec{x}=[x_1~x_2~...~x_N]^T$. In the case of two discrete random vectors, double subscripts $p$ and $k$ are used, where $p\in\{1,2\}$ is the vector index and $k$ is the element index, e.g., $X_{pk}$ denotes the $k$th element of the $p$th vector. Also note that, a continuous signal is always written with its dependency on another variable, e.g., $a(\lambda)$, whereas its discrete counterpart is often written in upper case with subscripts, e.g., $A_k$. At last, NFT and INFT operations are denoted as NFT$(\cdot)$ and INFT$(\cdot)$, $\mathcal{H}(\cdot)$ denotes the Hilbert transform.

\section{Review of NFT-INFT}\label{sec:NFT}
\begin{table}
   \centering
\caption{Fibre and system simulation parameters} \label{tab:fibre para}
\begin{tabular}{|c|c|c|}
  \hline

  $\nu$ & $193.44$ THz & centre carrier frequency \\
  \hline
  $\alpha$ & $0.2~\rm{dB}~ \rm{km}^{-1}$ & fibre loss\\
  \hline
  $\gamma$ & $1.3$ $\rm{(W\cdot km)}^{-1}$ & non-linearity parameter\\
  \hline
  $\beta_2$ & $-21.5\times 10^{-27}~\rm{s^2/m}$ & group velocity dispersion\\
  \hline
  $W$ & 56 GHz &linear bandwidth \\
  \hline
  $R_o$ & 8 & oversampling rate\\
  \hline
  $L_{\textnormal{sp}}$ & $80$ km & span length\\
  \hline
  $N_{\textnormal{sp}}$ & 12 & number of spans\\
  \hline
  NF & 5 dB & EDFA noise figure\\
  \hline
\end{tabular}
\end{table}%

We outline the NFT framework in the Manakov system. We refer the readers to \cite{Goossens17} for more details. 
To begin with,
the optical fibre channel model of concern is a multi-span point-to-point dual-polarisation dispersion unmanaged fibre link, which can be described by the Manakov equation \cite{AGRAWAL2013193}
\begin{align} 
\frac{\partial \vec{Q}}{\partial z}+\frac{\alpha}{2}\vec{Q}+\frac{j\beta_2}{2}\frac{\partial^2 \vec{Q}}{\partial t^2} -j\frac{8}{9}\gamma \vec{Q}\left\Vert\vec{Q}\right\Vert^2 = 0
\label{eq:manakov}
\end{align}
where $j=\sqrt{-1}$ and $\vec{Q}(t,z)= [Q_1(t,z)~Q_2(t,z)]$ is the complex envelope of the DP-signal as a function of time $t$ and distance $z$ along the fibre. The constants $\beta_2$, $\gamma$, and $\alpha$ are the chromatic dispersion, the Kerr non-linearity, and the loss coefficients of the fibre. At the end of each identical span of length  $L_{\rm sp}$, signals are amplified by an EDFA, which compensates the loss per span but adds  amplified spontaneous emission (ASE) noise. Tab.~\ref{tab:fibre para} lists all parameters used in this paper.

Because of the loss term, \eqref{eq:manakov} is not ``integrable'', which is a necessary condition to guarantee the linear evolution of the non-linear spectrum of signals propagating in the channel. The usual remedy is to approximate \eqref{eq:manakov} with the so-called path-averaged integrable model.
\begin{equation}
\frac{\partial \vec{Q_{\rm pa}}}{\partial z}+\frac{j\beta_2}{2}\frac{\partial^2 \vec{Q_{\rm pa}}}{\partial t^2} -j\frac{8}{9}\gamma_{a} \vec{Q_{\rm pa}}\left\Vert\vec{Q_{\rm pa}}\right\Vert^2 = 0
\label{eq:pa-model}
\end{equation}
where $\gamma_a=\gamma(1-e^{-\alpha L_{\rm sp}})/(\alpha L_{\rm sp})$. After each EDFA at $z = ML_{\rm sp},~M=1,2,...,N_{\rm sp}$, $\vec{Q_{\rm pa}}(t,z)$ \textit{approximates} $\vec{Q}(t,z)$ with only a small error. The detailed derivation can be found in \cite{Hasegawa:90, Le:15}. The approximation error will be discussed later in Sec.\ref{sec:nonintegrability}.

It is also convenient to normalise \eqref{eq:pa-model} with the factors
\[
T_n = \sqrt{(|\beta_2|Z_0)/2},~Q_n = \sqrt{2/(\frac{8}{9}\gamma_a Z_0)}  
\]
where $Z_0=N_{\rm sp}L_{\rm sp}$. With the normalised variables $\vec{q}(\ell,\tau) = \vec{Q_{\rm pa}}(z,t)/Q_n$, $\tau = t/T_n$ and $\ell = -z/Z_0$, \eqref{eq:pa-model} is simplified to 
\begin{equation}
j\frac{\partial}{\partial \ell}\vec{q}(\ell,\tau)=\frac{\partial^2 }{\partial \tau^2}\vec{q}(\ell,\tau)+2\left\Vert\vec{q}(\ell,\tau)\right\Vert^2 \vec{q}(\ell,\tau) 
\label{eq:normalisedManakov}
\end{equation}
where $\vec{q}(\ell,\tau) = [q_1(\ell, \tau)~q_2(\ell, \tau)]$. \eqref{eq:normalisedManakov} is integrable and its solution can be represented by NFT using the Zakharov-Shabat (ZS) system~\cite{Goossens17,Manakov1974}
\begin{equation*}
\frac{\partial \vec{v}(\ell,\tau, \lambda)}{\partial \tau} = \left(\begin{matrix}
-j\lambda & q_1(\ell,\tau) & q_2(\ell,\tau) \\
-q_1^*(\ell,\tau) & j\lambda & 0 \\
-q_2^*(\ell,\tau) & 0 & j\lambda \\
\end{matrix}\right)\vec{v}(\ell,\tau, \lambda)
\label{eq:ZS}
\end{equation*} 
where $\lambda$ is the non-linear frequency, $\vec{v} = [v_0~v_1~v_2]^T$ is the canonical eigenvector with the boundary condition 
\begin{equation*}
\vec{v}(\ell, \tau, \lambda) =\left( \begin{matrix}
1\\0\\0
\end{matrix}\right)e^{-j\lambda \tau},~\tau\rightarrow -\infty.
\label{eq:boundary}
\end{equation*}
We define the non-linear coefficients $a(\ell,\lambda)$, $b_1(\ell,\lambda)$ and $b_2(\ell,\lambda)$ as

\vspace{6pt}
{\centering
  $ \displaystyle
    \begin{aligned} 
a(\ell,\lambda) &= e^{+j\lambda\tau}v_0(\tau, \ell,\lambda),\\
b_1(\ell,\lambda) &= e^{-j\lambda\tau}v_1(\tau, \ell,\lambda),~~~~~\tau\rightarrow+\infty.\\
b_2(\ell,\lambda) &= e^{-j\lambda\tau}v_2(\tau, \ell,\lambda).\\
    \end{aligned}
  $ 
\par}
\vspace{8pt}
\noindent When $q_1(\ell,\tau)$ and $q_2(\ell,\tau)$ vanish faster than any exponential functions \cite[p.17]{ablowitz1981solitons}, which is the case of our signals, the above limits exist and well-defined. We refer the readers to \cite{ablowitz1981solitons} for the rigorous conditions on the signals for which the limits exist. 
Let define $\vec{b}(\ell,\lambda)=[b_1(\ell,\lambda)~b_2(\ell,\lambda)]$. 
The crucial property of the non-linear spectrum is 
\begin{align*}
a(\ell,\lambda) = a(0,\lambda),~~\vec{b}(\ell,\lambda) = e^{-4j\lambda^2\ell} \vec{b}(0,\lambda).
\end{align*}
The property indicates that the non-linear spectrum for each $\lambda$ evolves in $\ell$ independently by the above simple equation.
It is the main motivation to modulate information on non-linear frequencies. For the ease of presentation, we drop the dependence on $\ell$ when it is clear from the context.

The non-linear spectrum is usually partitioned in two parts: The continuous spectrum that includes the real-valued frequencies $\lambda\in \mathbb{R}$ and the discrete spectrum that includes the roots of $a(\lambda)$ on the upper complex plane, i.e. $\Omega=\{\lambda_m~|~a(\lambda_m) = 0, \lambda_m\in \mathbb{C}^+\}$. The roots are isolated points and called eigenvalues. Then the non-linear spectrum can be defined by the following ratios:
\begin{equation*}
\textnormal{NFT}(\vec{q}(\tau))=\begin{cases}
\dfrac{\vec{b}(\lambda)}{a(\lambda)},~~~~~\lambda\in \mathbb{R},\\[10pt]
\dfrac{\vec{b}(\lambda_m)}{a'(\lambda_m)},~~~\lambda_m\in \Omega,
\end{cases}
\end{equation*}
where $a'(\ell,\lambda)=\frac{\partial a(\ell,\lambda)}{\partial \lambda}$. This representation is commonly used, e.g. in \cite{ablowitz1981solitons,Yousefi2014,Goossens17}. We denote here the continuous spectrum by 
$\vec{q}_c(\lambda)=\vec{b}(\lambda)/a(\lambda)$.

Another more interesting representation of non-linear spectrum
is solely based on $\vec{b}(\lambda)$. It exploits the fact that $a(\lambda)$ can be expressed in terms of $\vec{b}(\lambda), \textnormal{for}~\lambda\in \mathbb{R}$ and $\vec{b}(\lambda_m)$ for $\lambda_m\in \Omega$.
While the general expression\footnote{The general expression is given for single-polarisation NFT, but the same lines of derivation work for the dual-polarisation NFT.} is given in \cite[p.~57, Eq.~(1.6.19)]{ablowitz1981solitons}\cite[p.~50, Eq.~(6.23)]{faddeev2007hamiltonian}, we state here its special case of our interest. In the absence of discrete spectrum,
\begin{equation}
\label{eq:a_from_b}
a(\lambda) = \sqrt{1-\norm{\vec{b}(\lambda)}^2}\exp\left(j\mathcal{H}\left(\log\sqrt{1-\norm{\vec{b}(\lambda)}^2}\right)\right).
\end{equation}
The above equation is the result of the \emph{unimodularity condition}\cite[Eq. (A.3)]{Manakov1974}
\begin{equation}
\bigl\lvert a(\lambda)\bigr\rvert^2+\bigl\lvert b_1(\lambda)\bigr\rvert^2+\bigl\lvert b_2(\lambda)\bigr\rvert^2 =1,~~~{\rm for }~~\lambda\in \mathbb{R}.
\label{eq:unimo}
\end{equation}
Clearly, one can then compute $\vec{q}_c(\lambda)$ from $\vec{b}(\lambda)$ and $a(\lambda)$.
The $b$-modulation scheme in \cite{wahls2017ecoc} for single polarisation is developed based on this latter representation.

In this paper, the discrete spectrum is not in use, and thus, the INFT is defined as a mapping from the continuous spectrum, either $\vec{q_c}(\lambda)$ or $\vec{b}(\lambda)$, to the time-domain signal $\vec{q}(\tau)$.

One of the most important parts of both NFT and INFT algorithms is numerically solving ZS system for the canonical eigenvector at $\tau=+\infty$ from the boundary condition at $\tau=-\infty$ or vice versa. 
Different algorithms exist, some of which are outlined in \cite{Turitsyn:17}.
For our implementation of DP-INFT/NFT algorithms, the readers are referred to \cite[Sec. 3]{Goossens17}, which is a straightforward extension of the inverse Ablowitz-Ladik scheme in \cite[Sec. V.B]{Yousefi2016arxiv} for single-polarisation.
It is also worth mentioning the Parseval identity in the non-linear frequency domain \cite[Eq. (1.7.8)]{ablowitz1981solitons}
\begin{align*}
\int \norm{\vec{q}(t)}^2 dt &= \frac{1}{\pi}\int \log\Bigl(1+\norm{\vec{q_c}(\lambda)}^2\Bigr) d\lambda\\
& = -\frac{1}{\pi}\int \log\Bigl(1-\norm{\vec{b}(\lambda)}^2\Bigr) d\lambda .
\end{align*}

\section{$q_c$- and $b$-modulated DP-NFDM Systems}\label{sec:qcb}

The independent evolution of $\vec{b}(\lambda)$ for each non-linear frequency $\lambda$ motivates
to modulate the data independently over separate sub-carriers in the non-linear spectrum as it is done by the orthogonal
frequency-division multiplexing (OFDM) schemes in the classical linear channels.
Therefore, we generate an NFDM symbol in the non-linear spectrum in a similar manner of an OFDM symbol in the (linear) Fourier spectrum. Here, we borrow the notations from \cite{Armstrong2009} to explain the generation of NFDM symbols, either in $b$-domain or in $q_c$-domain.

Consider the input and output of  the inverse discrete Fourier transform (IDFT).
The input of IDFT is a complex vector $\vec{x} = [ x_1~x_2~...~x_{N}]^T$ of length $N$. Each element $x_k$ represents the data on the corresponding sub-carrier, chosen from a given constellation, e.g., 32-QAM in this paper. The output of IDFT is a complex vector $\vec{y}$, representing the discrete time-domain samples. The IDFT is defined by
\begin{equation*}
y_m = \frac{1}{N}\sum_{k=0}^{N-1}x_k \exp(\frac{j2\pi km}{N}),~~{\rm for}~~0\leq m\leq N-1.
\end{equation*}
Correspondingly, the DFT is
\begin{equation*}
x_k = \sum_{m=0}^{N-1}y_m\exp(\frac{-j2\pi km}{N}),~~{\rm for}~~0\leq k\leq N-1.
\end{equation*}
Upon implementation, a suitable circular shift of vectors is assumed to make sure that the zero-frequency component of $\vec{y}$ is always at the centre of spectrum.
We also define $f(\vec{x},M)$ as a function that adds $M$ zeros to a vector $\vec{x}$ of length $N$ in the following way
\begin{equation*}
f(\vec{x},M) = [\underbrace{0~...~0}_{M/2}~ x_1~...~x_N~\underbrace{0~...~0}_{M/2}],
\end{equation*}
which is next used to describe the action of up-sampling and adding guard interval. 

The chromatic dispersion causes temporal broadening. 
Thus, some guard intervals (GI) between NFDM symbols are necessary
to avoid inter-symbol interference.
 The required interval $T_G$ can be estimated by \cite{le2018jlt}
\begin{equation}\label{eq:gi}
T_G \approx \pi W\beta_2 L_{\rm sp}N_{\rm sp}=3.75~\textnormal{ns},
\end{equation}
which depends mainly on signal bandwidth $W$ and transmission distance $L_{\rm sp}N_{\rm sp}$ that were both kept constant in this work (given in Tab.~\ref{tab:fibre para}). Let $N_C$ denote the number of sub-carriers in each polarisation. The pulse duration is $T_0=N_C/W$. The total pulse duration is $T_0+T_G$, which reduces the SE. We define $\eta = (T_0+T_G)/T_0$, showing the SE loss due to the guard interval.

\begin{figure}[tb!]
\centering
\begin{tabular}{c}
  \hspace*{-1.2cm}
  \begin{minipage}[!htp]{0.45\textwidth}
	\begin{tabular}{cc}
\includegraphics[width = 0.48\textwidth]{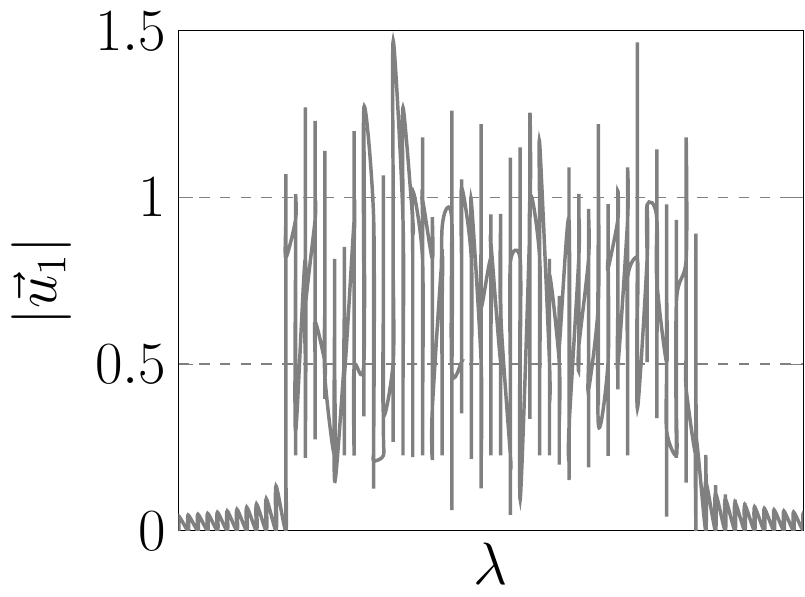}
	&
\includegraphics[width = 0.48\textwidth]{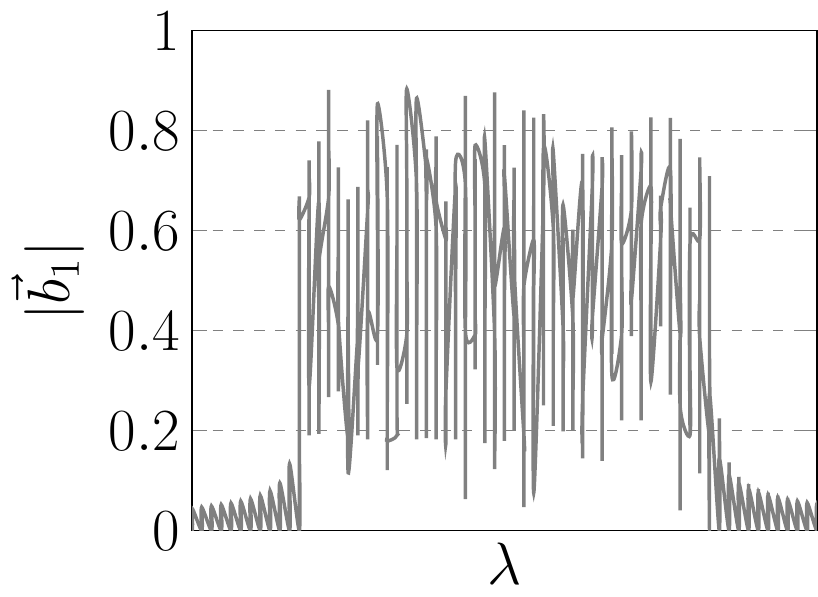}
    \\
     ~~~(a) & ~~~(b) \\
\includegraphics[width=0.48\textwidth]{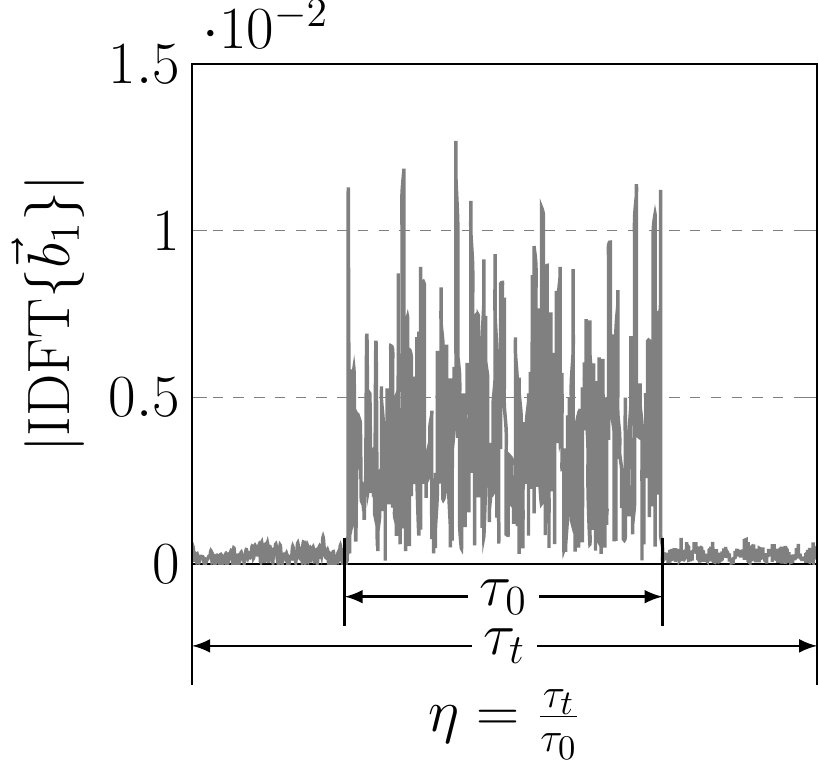}
    &
\includegraphics[width = 0.48\textwidth]{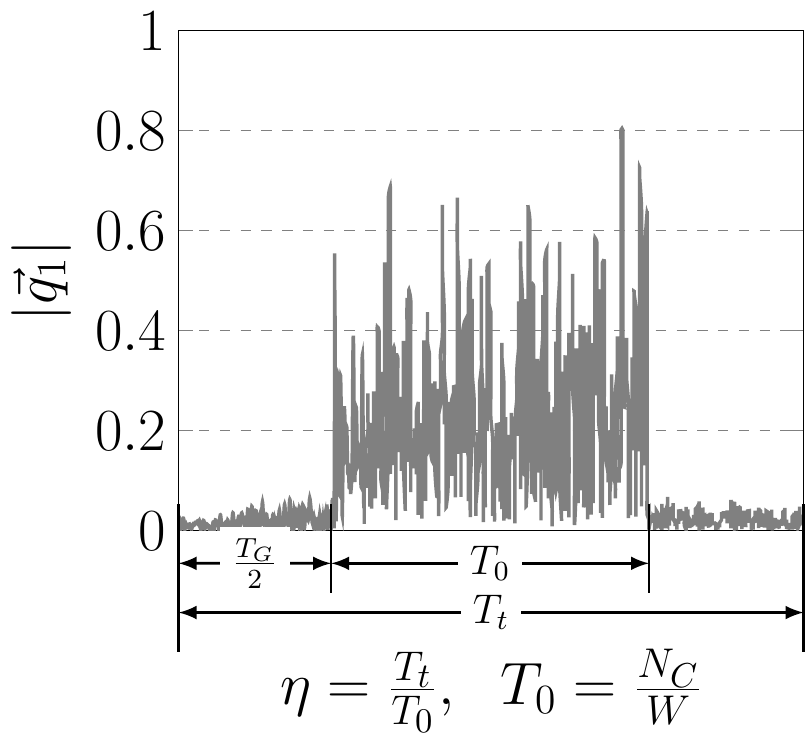}
    \\
  ~~~(c) & ~~~(d)\\
  	\end{tabular}	
  \end{minipage}\\
  \includegraphics[width = 0.43\textwidth]{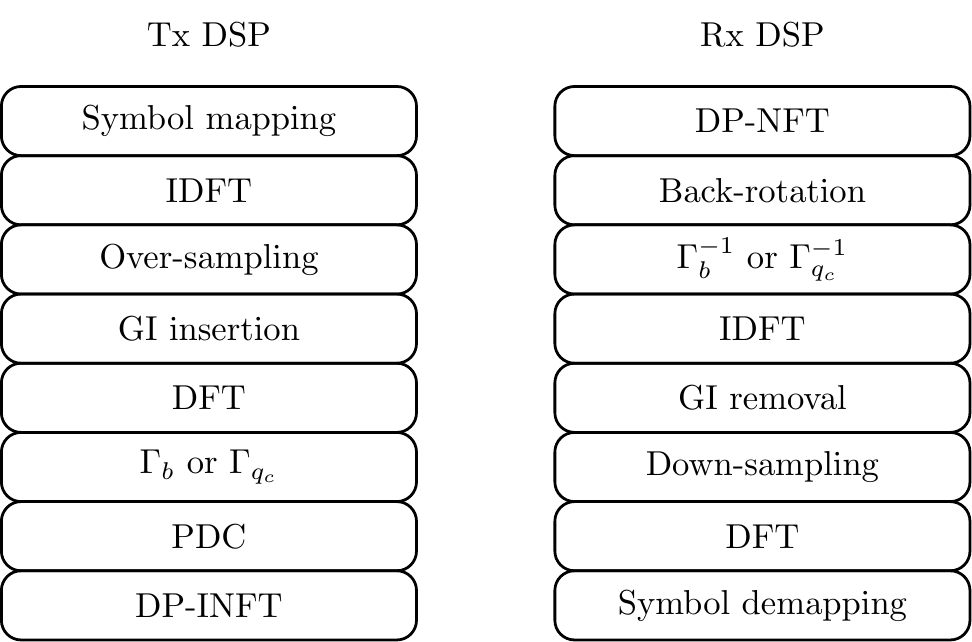}\\
  (e)
\end{tabular}
	\caption{A realisation of signal at different stages of INFT-signal generation without pre-dispersion compensation. (a) OFDM spectrum as $\vec{u}_1$, (b) $\vec{b}_1$ spectrum when applying the transformation $\Gamma_b$ on $\mathbf{u}$, (c) IDFT of $\vec{b}_1$ with tails due to the transformation $\Gamma_b$, (d) time-domain signal generated by INFT. (e) Transceiver digital signal processing structure.}
    \label{fig:signals}
\end{figure}

Let $\vec{x}_p=[x_{p1}~x_{p2}~...~x_{pN_C}]^T$ denote the data modulated on $N_C$ sub-carriers for 
each polarisation $p\in\{1,2\}$. Before computing the discrete samples 
of the continuous spectrum, we describe the action of adding GI and upsampling 
 \begin{equation*}
 \vec{d}_p=\textnormal{IDFT}\{f(\vec{x}_p,N_C(R_o-1))\},~~(\textnormal{up-sampling})
 \end{equation*}
 \begin{equation*}
  \vec{u}_p=\begin{cases}
\textnormal{DFT} \{f(\vec{d}_p,N_C R_o(\eta-1))\}, \eta\geq 2,\\
 \textnormal{DFT} \{f(\vec{d}_p,N_C R_o)\}, 1<\eta<2,\\
 \end{cases}~~(\textnormal{adding GI})
 \end{equation*}
where $R_o$ is the oversampling rate as in Tab.~\ref{tab:fibre para} and $N_C R_o(\eta-1)$ is rounded up to its nearest even number.
Next, the discrete samples of the continuous spectrum, in either $\vec{b}(\lambda)$ or $\vec{q_c}(\lambda)$, are computed from the resulted vectors $\vec{u}_p=[u_{p1}~u_{p2}~...~u_{pN}]^T,~p\in\{1,2\}$ 
as follows:
for $q_c$-modulation schemes, the samples of $\vec{q_c}(\lambda)=[q_{c1}(\lambda)~q_{c2}(\lambda)]$ are obtained from $[\vec{u}_1~\vec{u}_2]$ via the transformation
\begin{equation*}
  \begin{array}{rl}
q_{c,pk} &= \sqrt{e^{|u_{pk}|^2}-1}\cdot e^{j\arg\{u_{pk}\}},~ p\in\{1,2\}. 
 \end{array}
 \end{equation*} 
The above transformation, denoted here by $\Gamma_{c}$, is already explored for DP-NFDM systems in \cite{Goossens17}. 
For $b$-modulation schemes, the samples of $\vec{b}(\lambda)=[b_1(\lambda)~b_2(\lambda)]$ are obtained from $[\vec{u}_1~\vec{u}_2]$ via a new transformation
\begin{align*}
\Delta_k & =\sqrt{ 1-\exp\Bigl(-|u_{1k}|^2-|u_{2k}|^2\Bigr)}/\Bigl(|u_{1k}|^2+|u_{2k}|^2\Bigr),\\
b_{1k} & = \Delta_k\cdot u_{1k},\\
b_{2k} & = \Delta_k\cdot u_{2k},\\
a_k & = \sqrt{1-|b_{1k}|^2-|b_{2k}|^2}e^{j\mathcal{H}\bigl(\frac{1}{2}\log(1-|b_{1k}|^2-|b_{2k}|^2)\bigr)}.
\end{align*}
The above transformation, denoted here by  $\Gamma_b$, is the natural extension of the one introduced in \cite{yousefi2016ecoc} for SP-NFDM systems in the defocusing regime.
Before taking INFT, the technique of pre-dispersion compensation (PDC) is applied on the resulted continuous spectrum, either $\vec{b}(\lambda)$ or $\vec{q_c}(\lambda)$, to minimise the required GI\cite{Tavakkolnia2016}. Finally, the time-domain NFDM signal of duration $T_0+T_G$ is generated by applying DP-INFT algorithm. At the receiver, the DP-NFT and back-rotation equalisation are performed. Fig.~\ref{fig:signals} illustrates the transceiver digital signal processing structure and a realisation of signals at different stages of INFT-signal generation are illustrated. Note that, for $\eta <2$, each signal generated by INFT has a duration of $2T_0$. Before transmission, the signal will be truncated symmetrically on both sides to a signal of duration $T_0+T_G$ for transmission and also recovered to the duration $2T_0$ by adding zeros before NFT processing. This is due to the heuristic reason that higher spectral resolution is needed for NFT as $N_C$ increases.

\begin{figure*}[htp!]
   \centering
    \includegraphics[width=0.78\textwidth]{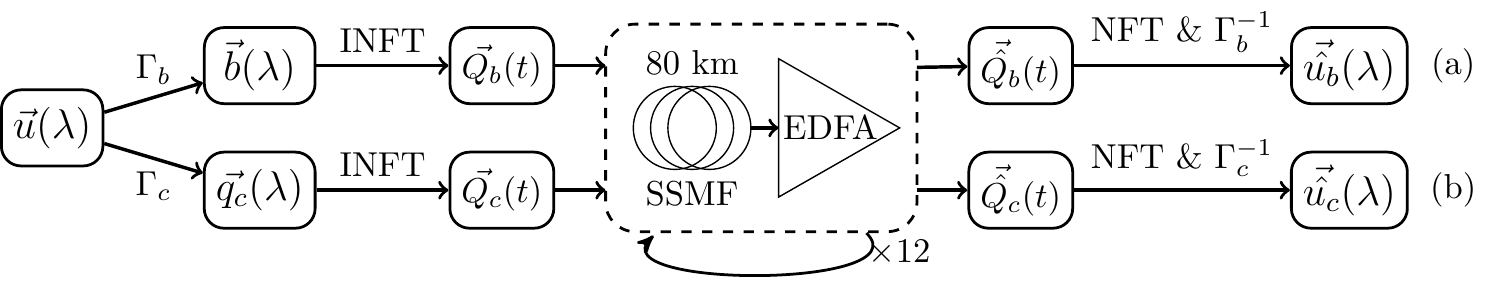} 
    \caption{ Simulation diagram of (a) $b$- or (b) $q_c$-modulated DP-NFDM systems. $\vec{Q_b}(t)$ and $\vec{Q_c}(t)$ are the time-domain signals of each system.}
    \label{fig:entropy_sym}
\end{figure*}

For any $\lambda\in\mathbb{R}$, the constraint $|b_1(\lambda)|^2+|b_2(\lambda)|^2<1$ specifies a 4-dimensional unit ball. The transformation $\Gamma_b$ facilitates the signal modulation on $b$-domain as it maps the entire $\mathbb{R}^4$ into this unit ball. Therefore, it allows using any arbitrary constellation format, e.g. QAM or equi-distance ring constellations without any concern on violating the constraint on $b$-coefficients. 
However, the transformation $\Gamma_b$ has a drawback. As it is shown in \cite{wahls2017ecoc},
modulating directly $b$-coefficients allows a full control on the duration of NFDM signals in time-domain.
Applying non-linear transformations like $\Gamma_b$ causes losing this property to some extent. 
Nevertheless, this may not be practically a big issue.
Although the generated NFDM signal may have some undesired tails, the tails are partially covered by the necessary long GI (specially after PDC) when the fibre length is long enough. Note that the longer GI is, the more ASE noise
is taken into the NFT process, causing Q-factor degradation. The optimal GI that balances the extra ASE noise and the truncation of decaying tails is close to the estimated $T_G$ in \eqref{eq:gi}.

Directly modulating $b$-coefficients of SP-NFDM was investigated in \cite{Le:18,Gui:18}. 
In \cite{Le:18}, the clipping technique was applied to satisfy the constraint on $b(\lambda)$. Clipping, an non-invertible operation commonly used for the peak-to-average power ratio (PAPR) reduction in OFDM systems \cite[Sec. V-A]{Armstrong2009}, inevitably introduces distortion in $b(\tau)$ and cannot lift the constraint completely. 
In \cite{Gui:18}, another alternative was proposed to guarantee the constraint on $b$-coefficients by optimising the carrier waveform and applying constellation shaping. The technique was applied for 9 sub-carriers. It is of practical interest how the design complexity and the achievable SE scale for large number of sub-carriers. Moreover, the above techniques were presented in single polarisation. It is of interest how 
the joint constraint on $b_1(\lambda)$ and $b_2(\lambda)$ in dual polarisation, which limits independent modulation of $b_1(\lambda)$ and $b_2(\lambda)$, may affect the application of both techniques.

\begin{figure}[!tb]
\centering
   \includegraphics[scale = 0.6]{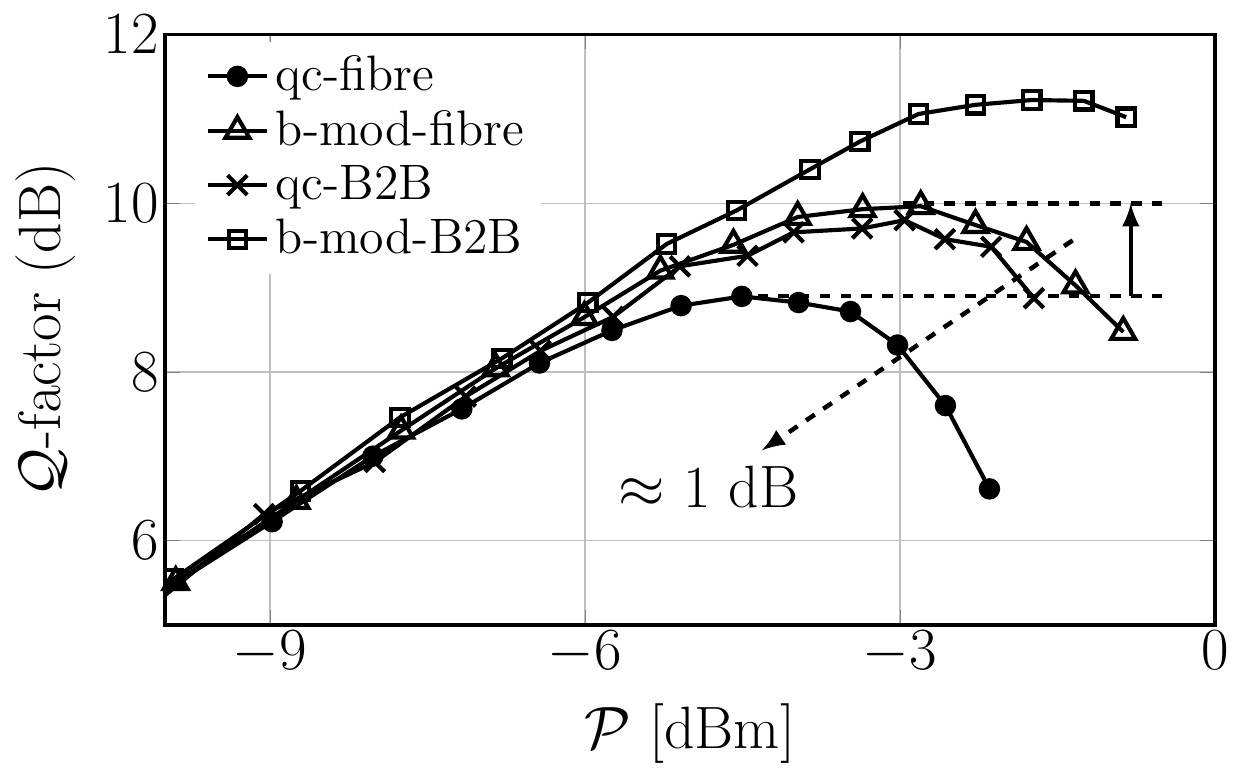}
   \caption{Q-factor vs. signal launch power in $q_c$- and $b$-modulated DP-NFDM systems with a large guard interval $\eta=4$ in both AWGN case and fibre transmission. The total additive noise powers are the same.}
   \label{fig:Qb2b}
\end{figure}

We demonstrate now the advantage of $b$-modulation over $q_c$-modulation in a relatively ideal case having a large guard interval $\eta = 4~(N_C = 70)$.
The transmission of corresponding NFDM symbols are compared in two different simulated transmission 
scenarios: 1) the transmission over 12 spans of 80 km SSMF with EDFA amplification as illustrated in Fig.~\ref{fig:entropy_sym}. It was simulated by the split-step Fourier method (SSFM) \cite{Sinkin2003} using the parameters in Tab.~\ref{tab:fibre para}. 2) the additive white Gaussian noise (AWGN) channel without fibre transmission (back-to-back). The total additive noise powers were the same in both scenarios. 
Fig.~\ref{fig:Qb2b} shows the clear advantage of $b$-modulation over $q_c$-modulation in both scenarios. Extra Q-factor degradation in the fibre transmission is attributed to the approximation error of the path-averaged model. The power $\mathcal{P}$  reported in this work is always the average power per polarisation.

\section{Optimisation of $b$-modulated DP-NFDM Transmission}\label{sec:optim}

\begin{figure}[tp!]
\centering
	\begin{tabular}{c}
  \hspace*{-0.9cm}  
    \includegraphics[scale = 0.6]{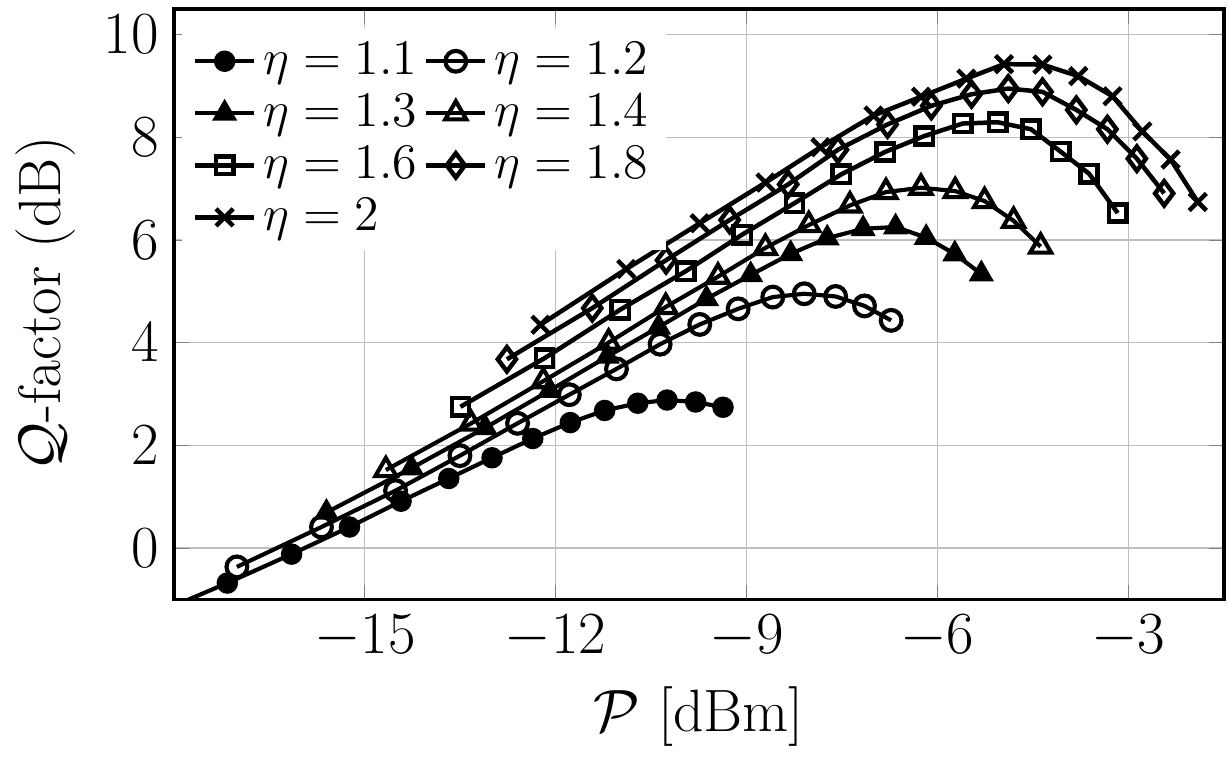}\\
(a)\\
  \hspace*{-1cm} 
   \includegraphics[scale = 0.6]{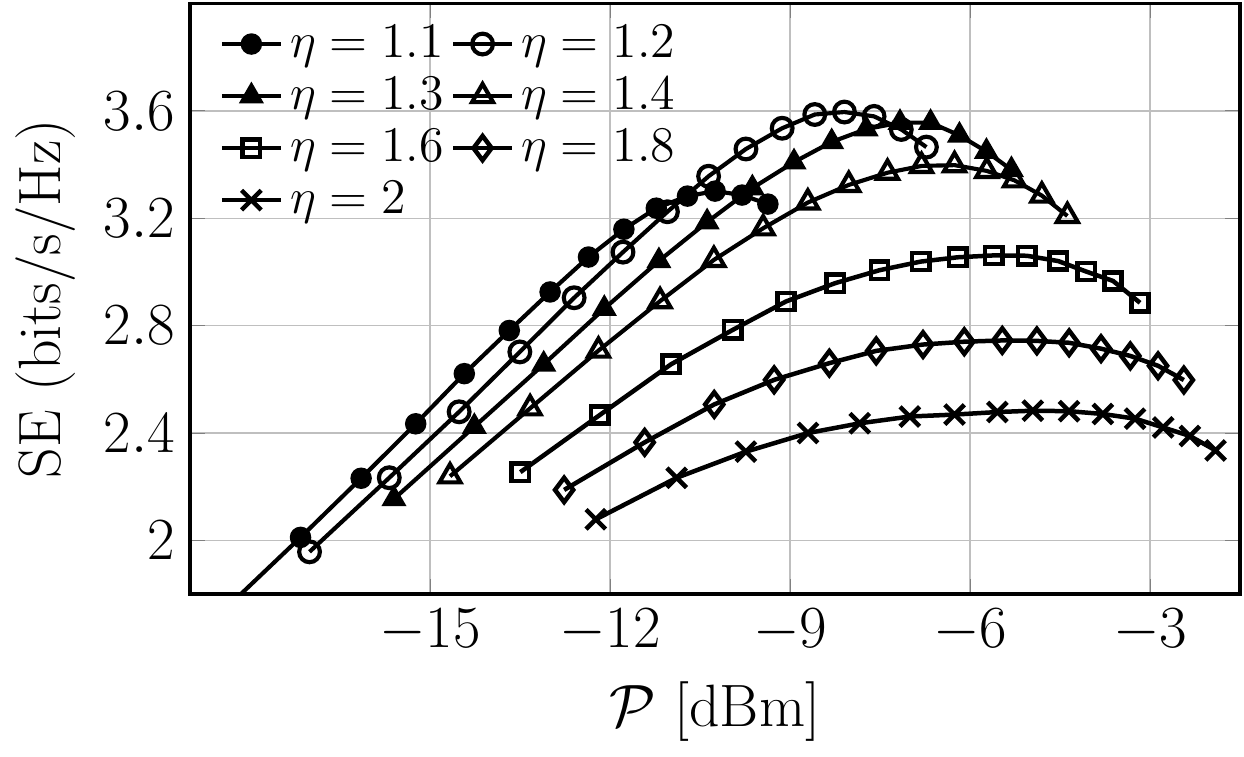}\\
  (b)\\
   \begin{minipage}[!htp]{0.46\textwidth}
     \centering
     \begin{tabular}{cc}
      \includegraphics[width = 0.35\textwidth]{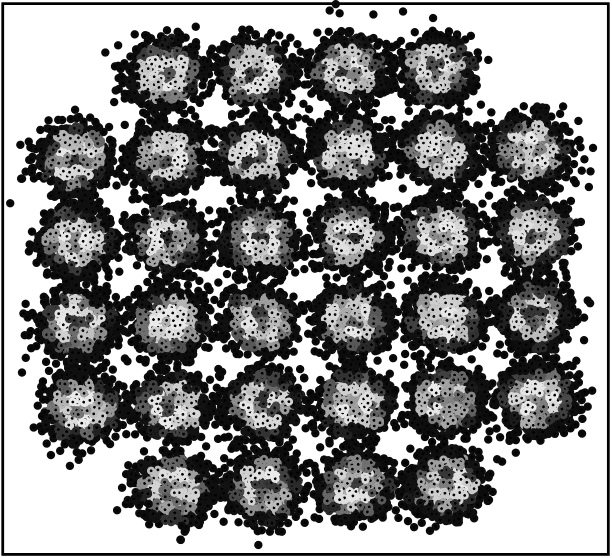} &
     \includegraphics[width = 0.35\textwidth]{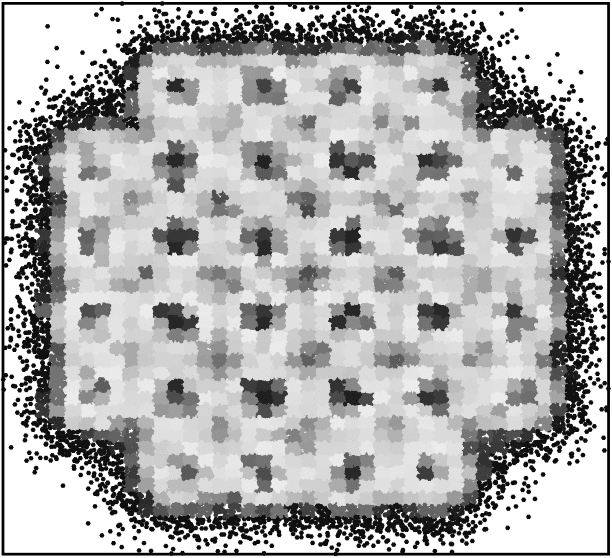} \\
   (c) & (d)  
   \end{tabular}    
   \end{minipage}\\
	\end{tabular}	
	\caption{The impact  of decreasing $\eta$ on (a) Q-factor and (b) SE in $b$-modulated DP-NFDM system  for different launch powers. $N_C=210/(\eta-1)$. The corresponding received constellation at the power of the maximum Q-factor (c) and SE (d).}
    \label{fig:etaopt}
\end{figure}

\begin{figure}
\centering
   \hspace*{-1cm} 
   \includegraphics[scale = 0.6]{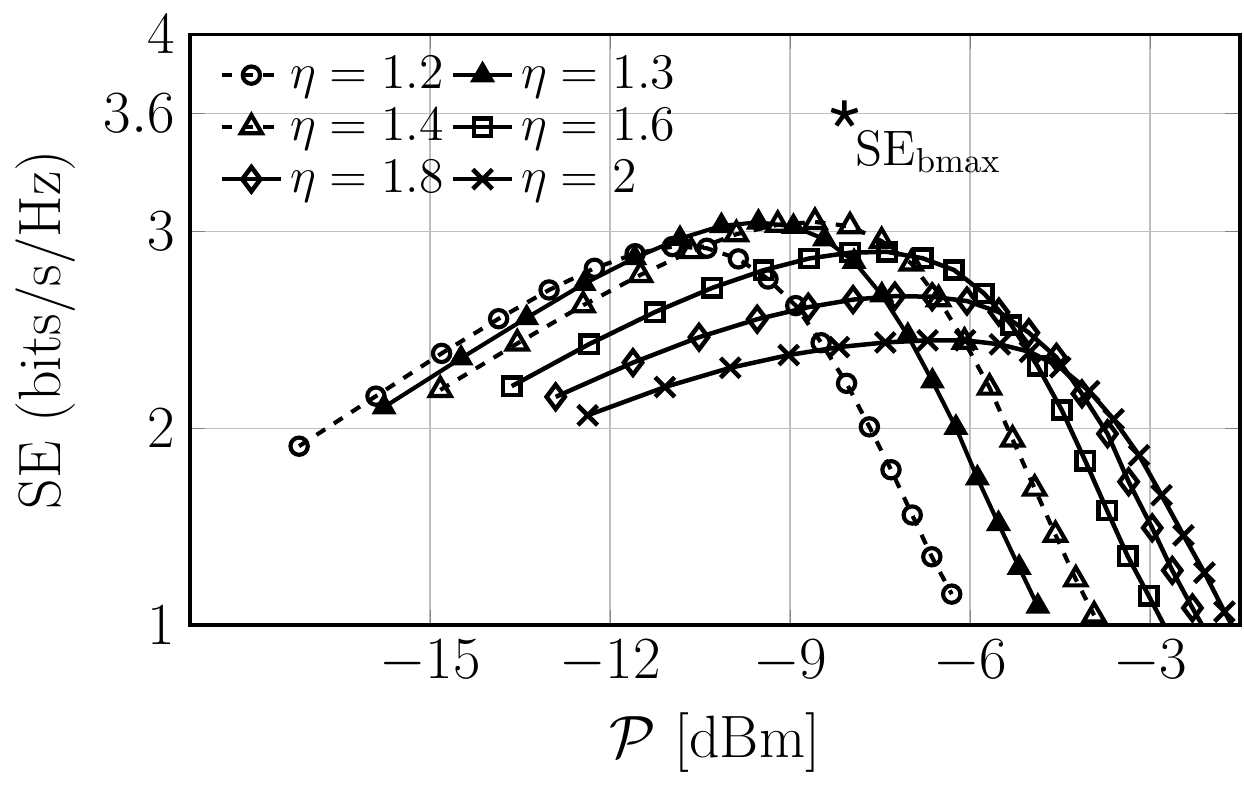}
   \caption{The impact  of decreasing $\eta$ on SE in $q_c$-modulated DP-NFDM system  for different launch powers. $N_C=210/(\eta-1)$.}
   \label{fig:qcSE}
\end{figure}

Once the advantage of $b$-modulation is established, we optimise the $b$-modulated DP-NFDM system in terms of net data rate 
for the transmission scenario over $12\times80$ km SSMF with EDFAs, visualised in Fig.~\ref{fig:entropy_sym}. The optimisation parameters are the ratio $\eta$ and launch power per polarisation $\mathcal{P}$. The number of sub-carriers $N_C$ and $\eta$ are connected by 
\begin{equation*}
\eta = (T_0+T_G)/T_0=1+WT_G/N_C,
\end{equation*}
where $T_G=3.75$~ns and $W$ is fixed.
For instance, it results $N_C=210$ for $\eta=2$. 
For each pair of $(\eta, \mathcal{P})$, the transmission of the corresponding NFDM symbols was simulated according to the parameters in Tab.~\ref{tab:fibre para}.
For each NFDM symbol, the sub-carriers were detected individually and the performance was averaged over a large number of randomly generated NFDM symbols.
To compute the SE, we estimate the mutual information (MI) (of individual sub-carrier detection) by assuming the channel conditional distribution as a Gaussian distribution~\cite[Sec. VI]{Arnold2006}\cite{Alvarado:18}.
This assumption underestimates the correct MI and serves as an achievable information rate. The SE is then computed by
\begin{equation*}
    \textnormal{SE} = \frac{\textnormal{MI}\cdot N_C}{(T_0+T_G)W}=\textnormal{MI}/\eta~(\rm bits/s/Hz/Pol.).
\end{equation*}
The computation of SE is to be taken in caution as we do not consider the spectral broadening of the signal in the fibre channel.
We expect that decreasing $\eta$ will increase SE as well as the signal-noise interaction for a given launch power $\mathcal{P}$. For different values of $\eta$, the transmission performance metric Q-factor is plotted in Fig.~\ref{fig:etaopt}(a) and SE in Fig.~\ref{fig:etaopt}(b). They show respectively that Q-factor decreases monotonically as $\eta$ decreases, while SE reaches its maximum at $\eta=1.2$.
Two received constellations at the power of the maximum Q-factor and SE are also plotted in Fig.~\ref{fig:etaopt}(c) and \ref{fig:etaopt}(d).
For $\eta=1.2$ ($N_C=1050$), and at power $\mathcal{P}=-8$ dBm per polarisation,
we achieved a SE of $3.6$ bits/s/Hz/pol, resulting $400$ Gbps net data rate, while the gross data rate is $560$ Gbps. 
For the sake of comparison, we repeat the simulation for $q_c$-modulated DP-NFDM systems and plotted the SE curves in Fig.~\ref{fig:qcSE}. 

\section{Co-variance Matrix of $q_c$- and $b$-modulated signals}\label{sec:cov}

In this section, we give some insights why the $b$-modulation scheme 
outperforms the $q_c$-modulation scheme. Let us first justify verbally.
As we mentioned in Sec.~\ref{sec:NFT}, there is a one-to-one correspondence between $\vec{b}(\lambda)$ and 
$\vec{q}_c(\lambda)$ provided that  
$\vec{b}(\lambda)~\textnormal{or} ~\vec{q}_c(\lambda)~\forall \lambda\in\mathbb{R}$ and all discrete
eigenvalues $\{\lambda_m\}$ are known. Therefore,   
$\vec{b}(\lambda)~\forall\lambda\in\mathbb{R}$ carries the same amount of information as $\vec{q}_c(\lambda)$ if the correlation
between spectral coefficients, in either $\vec{b}(\lambda)$ or $\vec{q_c(\lambda)}~\forall\lambda\in\mathbb{R}$, can be fully exploited and no discrete eigenvalue emerges\footnote{The discrete eigenvalues may emerge because of non-ideal amplification or mixing with ASE noise. However, we did not observe any significant sign of their presence in our simulations.}.
However, our two DP-NFDM schemes have different performance for the following reasons:
\begin{itemize}
\item[$(i)$] The sub-carriers are detected individually, where
the correlation between sub-carriers is neglected. Strong correlations between sub-carriers
cause then a large performance degradation.
\item[$(ii)$] The two DP-NFDM schemes use the same modulation technique, i.e. OFDM with 32-QAM modulated sub-carriers.
Depending on the distribution of noisy received data, this particular modulation format can be more resilient against noise for one scheme than the other.
\end{itemize}

We focus here on the first item, the correlation between sub-carriers. 
We show that the sub-carriers of $b$-modulated DP-NFDM system have much less 
correlation than the ones of $q_c$-modulated DP-NFDM system. 
To quantify all correlation coefficients with a single meaningful scalar, we use the differential entropy.
We compare two quantities: the joint entropy which takes the correlations into account and the individual entropy which neglects the correlations. The gap between these quantities reflects how much the sub-carriers are correlated and how much the net data rate will be decreased (to some extent) by neglecting the correlations. 

\begin{figure}[htp!]
\centering
	\begin{tabular}{c}
  \hspace*{-0.7cm}  
    \includegraphics[scale = 0.6]{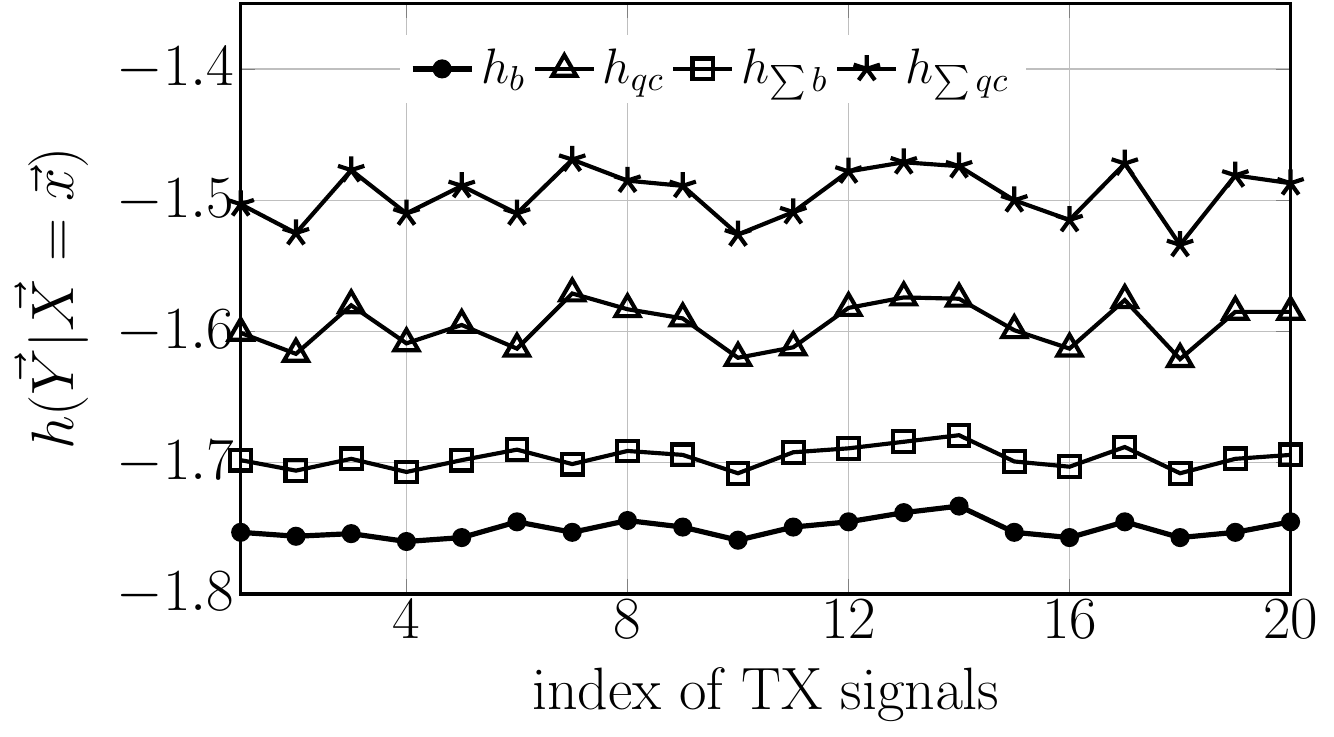}\\
(a)\\
  \hspace*{-0.7cm} 
   \includegraphics[scale = 0.6]{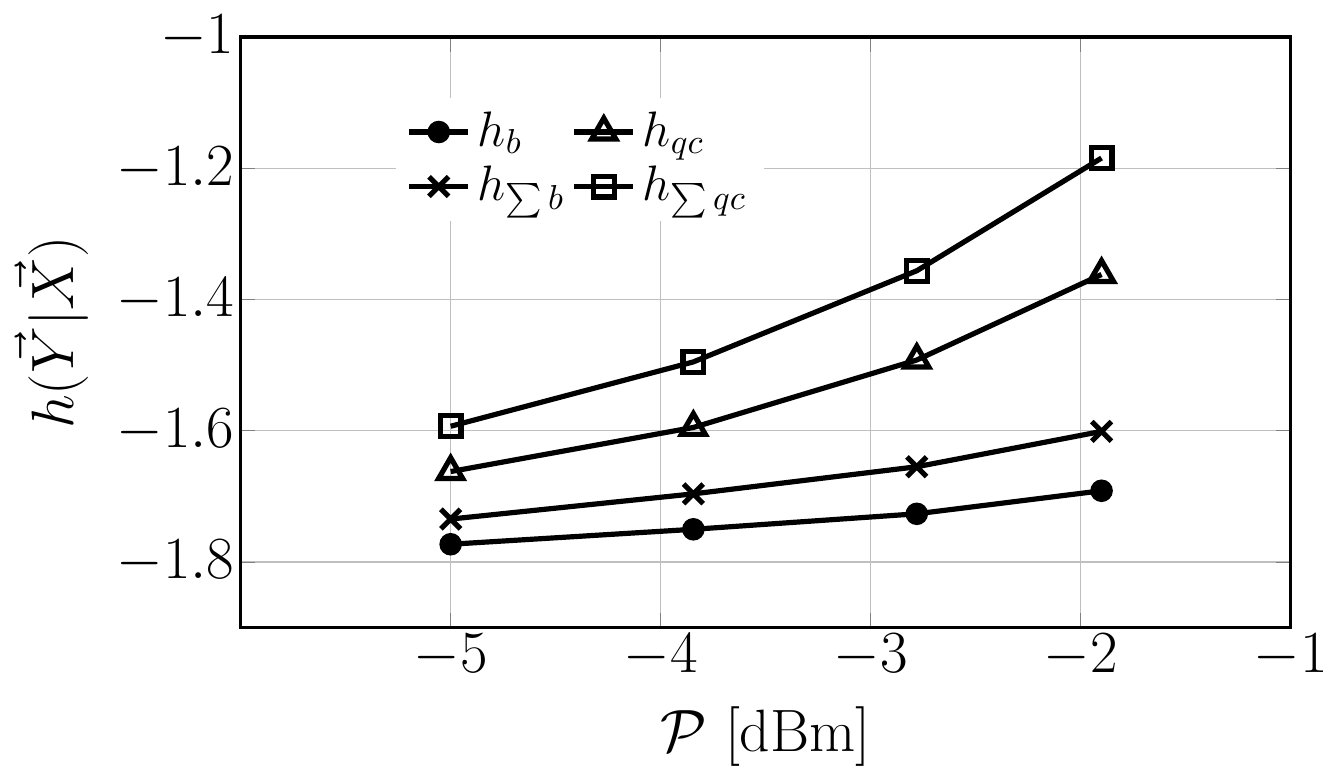}\\
  (b)\\
	\end{tabular}	
	\caption{Conditional differential entropy of received symbols in both $q_c$ and $b$-modulated DP-NFDM systems with the assumption of Gaussian distribution, for different transmitted symbols (a) and powers (b).}
	\label{fig:entropies}
\end{figure}

For both DP-NFDM systems, we numerically compute the differential entropies. Let $\vec{X}=[X_1~X_2~...~X_{2N_C}]$ denote 
the randomly chosen 
32-QAM data modulated on $2N_C$ sub-carriers (DP). Let $\vec{Y}=[Y_1~Y_2~...~Y_{2N_C}]$ denote the received noisy data of sub-carriers after the NFT processing. The joint entropy is defined as
\begin{align*}
    h(\vec{Y}\vert \vec{X})&=\frac{1}{4N_C}\mathbb{E}_{\vec{x}}\left(h(\vec{Y}\vert \vec{X}=\vec{x})\right)
\end{align*}
where $h(\vec{Y}\vert \vec{X}=\vec{x})$ is the entropy of received data given $\vec{x}$, a realisation of inputs $\vec{X}$. $\mathbb{E}_{\vec{x}}$ denotes taking expectation over all realisations of $\vec{x}$. The individual entropy is defined as
\begin{align*}
    h_\Sigma(\vec{Y}\vert \vec{X})&=\frac{1}{4N_C}\mathbb{E}_{\vec{x}}\left(\sum_{i=1}^{2N_C}h(Y_i\vert \vec{X}=\vec{x})\right)
\end{align*}
where $h(\vec{Y_i}\vert \vec{X}=\vec{x})$ is the entropy of the individual received sub-carrier $Y_i$
given the realisation $\vec{x}$. To compute the above quantities, we need the conditional probability density $f({\vec{Y}\vert \vec{X}})$. It is, however, unknown yet for NFDM systems and difficult to estimate due to its large dimension $4N_C$.
%
%
We approximate $f({\vec{Y}\vert \vec{X}})$ by a complex multivariate  Gaussian distribution. We justify our choice later in \textbf{Remark~1.}
Define $\vec{W}=[\text{Re}(\vec{Y}),\text{Im}(\vec{Y})]$, then the corresponding probability density $f({\vec{W}\vert \vec{X}})$ is approximated by a real-valued multivariate  Gaussian distribution with covairance matrix $\mathbf{K}$. The element of $\mathbf{K}$ is defined as 
\begin{equation*}
k_{ij} = \mathbf{E}[ (W_i-\mathbf{E}[W_i])(W_j-\mathbf{E}[W_j])].
\end{equation*}
In this case, the conditional entropies can be approximated as
\begin{align}
h(\vec{Y}\vert \vec{X}=\vec{x})&\approx\frac{1}{8N_C}\log\left(\det\left( 2\pi e \mathbf{K}\right)\right),\\
h_\Sigma(\vec{Y}\vert \vec{X}=\vec{x})&\approx \frac{1}{8N_C}\log( \prod_{i=1}^{2N_C} 2\pi e k_{ii}).
\end{align}
We estimated $\mathbf{K}$ for each realisation $\vec{x}$ from simulated data. We simulated the fibre transmission scenario visualised in Fig.~\ref{fig:entropy_sym} for both DP-NFDM systems. We set $\eta=2$, resulting $N_C=210$ sub-carriers. For each DP-NFDM system, we randomly generated 20 input realisations $\vec{x}$. For each realisation, the transmission of its corresponding DP-NFDM symbol was simulated $2^{14}$ times with different random noise realisations. The simulation is repeated for different launch powers $\mathcal{P}$.
Accordingly, the above entropies are computed from the resulted $\mathbf{K}$ for each input realisation $\vec{x}$. Fig.~\ref{fig:entropies}(a) shows the entropies of each realisation $\vec{x}$
for both DP-NFDM systems at launch power $\mathcal{P}=-3.75$ dBm. We observe that the curves have small variations for different realisations $\vec{x}$. 
We finally estimated $ h(\vec{Y}\vert \vec{X})$ and $h_\Sigma(\vec{Y}\vert \vec{X})$ by the empirical average of 20 input realisations. 

Fig.~\ref{fig:entropies}(b) illustrates the conditional entropies in terms of launch power for both DP-NFDM systems. We observe that the gap $h(\vec{Y}\vert \vec{X})-h_\Sigma(\vec{Y}\vert \vec{X})$ is smaller for the $b$-modulated scheme than the one for the $q_c$-modulated scheme. It implies that the penalty of individual detection is smaller for the $b$-modulated scheme. For both schemes, the increase of the gap in terms of launch power implies that the sub-carriers becomes more correlated for larger $\mathcal{P}$.  
Moreover, $h(\vec{Y}\vert \vec{X})$ of the $b$-modulated scheme is not only smaller but also has a slower growth in terms of $\mathcal{P}$. 
It indicates that in comparison of both systems, the co-variance matrix $\mathbf{K}$ of the $b$-modulated scheme has, on average, smaller diagonal elements which grow also slower in $\mathcal{P}$.
In plain language, the ``effective noise'' contaminating $b$-modulated scheme has smaller power and its power grows also slower with increasing $\mathcal{P}$ than the one of $q_c$-modulation. Let us emphasise that the constellation of both DP-NFDM systems are the same in $u(\lambda)$ domain (shown in Fig.~\ref{fig:entropy_sym}) for a fair comparison in Fig.~\ref{fig:entropies}(b).

\begin{figure}[h]
\centering
\includegraphics[scale = 0.6]{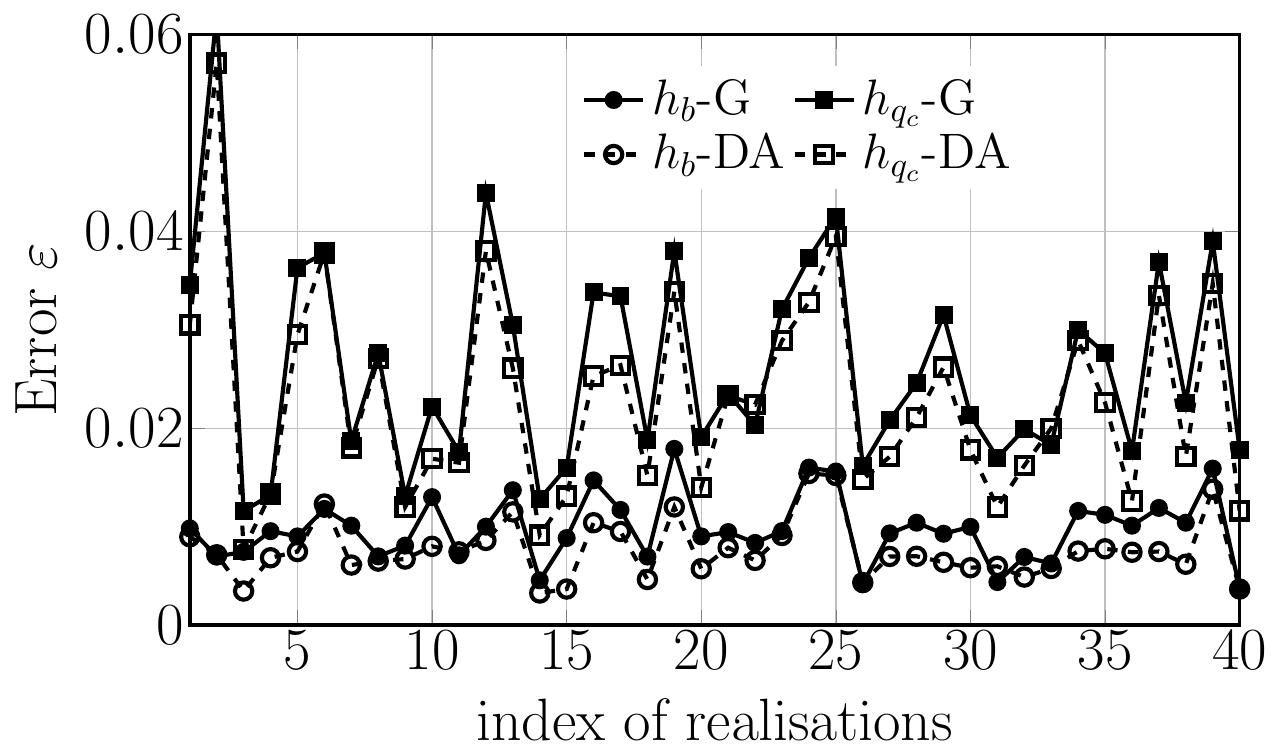}
\caption{The relative error between distribution-agnostic entropy estimator \cite{Lombardi16} and the Gaussian entropy for 10 adjacent sub-carriers. G stands for Gaussian entropy, DA for distribution-agnostic entropy estimation. 
For both methods, the result of one-dimensional entropy $h(Y_i|\vec{x})$ is very close. $\varepsilon=\big(\sum_{i=1}^{10} h(Y_i|\vec{x})-h(Y_{1},...,Y_{10}|\vec{x})\big)/\sum_{i=1}^{10} h(Y_i|\vec{x})$}.
\label{fig:GDAen}
\end{figure}

\noindent\textbf{Remark~1.} To verify the validity of entropy approximations, we further applied the entropy estimator in \cite{Lombardi16}, which can rather precisely estimate 
the differential entropy of a low-dimensional random variable from its random samples. To have robust estimations, we estimated the differential entropies of only 10
adjacent sub-carriers (20 real-valued dimensions). For different input realisation $\vec{x}$,
we observed that the estimations are fairly close to the corresponding entropies of the Gaussian approximation, as shown in Fig.~\ref{fig:GDAen}.

\noindent\textbf{Remark~2.} Comparing both DP-NFDM systems in terms of correlation and conditional entropy is more qualitative than quantitative. The joint MI between transmitted and received data on all sub-carriers 
is the suitable metric to ``quantify'' the performance difference of both NFDM schemes (with joint detection). The MIs using individual detection are already compared in Sec.~\ref{sec:optim}. However, the joint MI requires $f(\vec{Y}\vert\vec{X})$ as well as the probability density $f(\vec{Y})$.
It is more challenging to estimate $f(\vec{Y})$ specially when the input $\vec{X}$ is drawn from a discrete constellation, e.g., 32-QAM in our case. The probability density $f(\vec{Y})$ is a multi-dimensional multi-center distribution, with correlation between dimensions. The entropy $H(\vec{Y})$ is nearly impossible to calculate.
Hence, this problem makes the estimation of MI (jointly between all sub-carriers) much more challenging.

\section{Some Drawbacks of the Current System}\label{sec:drawbacks}
Based on the above simulation results, we identified two main sources of error in our current system: `transceiver distortion' in noiseless back-to-back scenario and channel mismatch owing to its non-integrability. 

\subsection{Transceiver Distortion in Noiseless Back-to-Back Scenario}\label{sec:truncation}
\begin{figure}[!h]
\centering
\includegraphics[scale = 0.6]{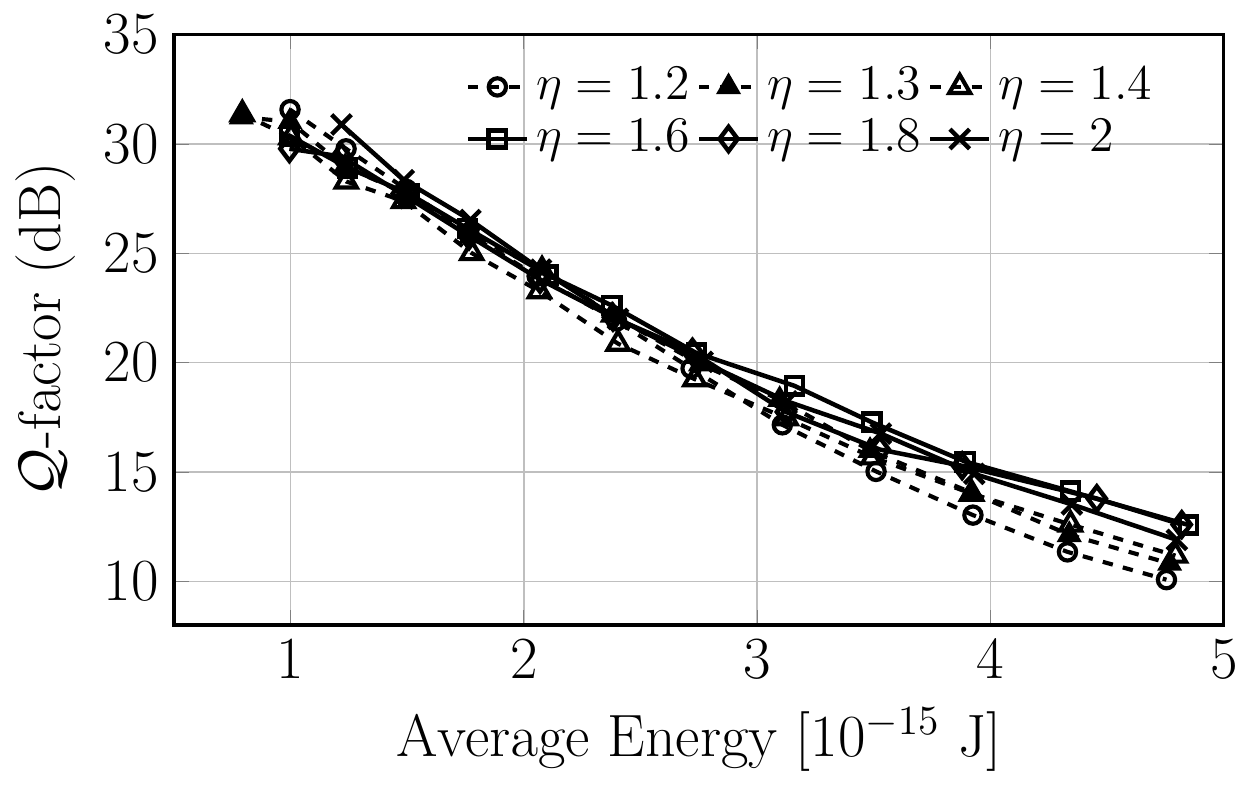}
\caption{ `Transceiver distortion' in the $b$-modulated DP-NFDM system in the noiseless back-to-back scenario. The average energy of NFDM symbol is calculated by $\mathcal{P}(T_0+T_G)=\mathcal{P}\eta N_C/W$.}
\label{fig:trunc}
\end{figure}

The transceiver distortion includes the distortion from the whole digital signal processing shown in Fig.~\ref{fig:signals}(e), such as the inaccuracy of the INFT-NFT algorithm, the application of $\Gamma_b$ and the truncation of time-domain signal.  Their overall effect can be easily quantified in an noiseless back-to-back simulation of the $b$-modulated DP-NFDM system. Fig.~\ref{fig:trunc} shows the Q-factor in terms of average energy per NFDM symbol for different $\eta$\correctiondelete{ in }. The average energy of one NFDM symbol can be calculated by $\mathcal{P}(T_0+T_G)=\mathcal{P}\eta N_C/W$. It shows clearly that the transceiver distortion is energy-dependent.

\subsection{Channel Mismatch due to Non-integrability}\label{sec:nonintegrability}
NFT is only exact for integrable channels such as \eqref{eq:pa-model}. To successfully apply NFT to the non-integrable channel \eqref{eq:manakov}, an approximation step has to be taken. To quantify the approximation distortion, one should simulate noiseless systems in Fig.~\ref{fig:entropy_sym} with a transceiver DSP that does not cause any distortion. This is unfeasible with our current transceiver DSP. To adopt a different approach, we first clarify the concepts of different digital back-propagation (DBP) schemes, as DBP is the main tool in our error quantification. 

We recall the Manakov equation \eqref{eq:manakov}. The ideal DBP refers to the process of solving the Manakov equation to recover $\vec{Q}(0,t)$ (input) from the boundary condition $\vec{Q}(z,t)$ (output), using SSFM with fine step size (0.1 km). The EDFAs are replaced with attenuators of the opposite gain. In a single-channel scenario, the ideal DBP should fully cancel the deterministic distortion.

We recall also the integrable path-averaged Manakov equation \eqref{eq:pa-model}. The path-averaged DBP refers to the process of solving the path-averaged Manakov equation to recover $\vec{Q}(0,t)$ from the boundary condition $\vec{Q}(z,t)$, using SSFM with fine step size (0.1 km). Since the signal propagation occurs in the Manakov equation, applying path-averaged DBP inevitably causes distortion on the recovered signal $\vec{Q}(0,t)$ even in the single-channel scenario. 

\begin{figure}
   \centering
      \begin{tabular}{c}
\includegraphics[width=0.4\textwidth]{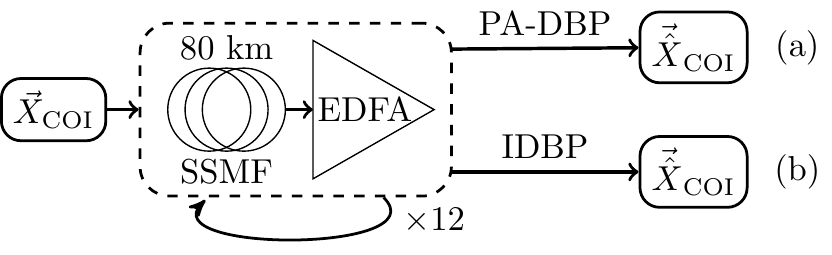}\\
\includegraphics[scale = 0.6]{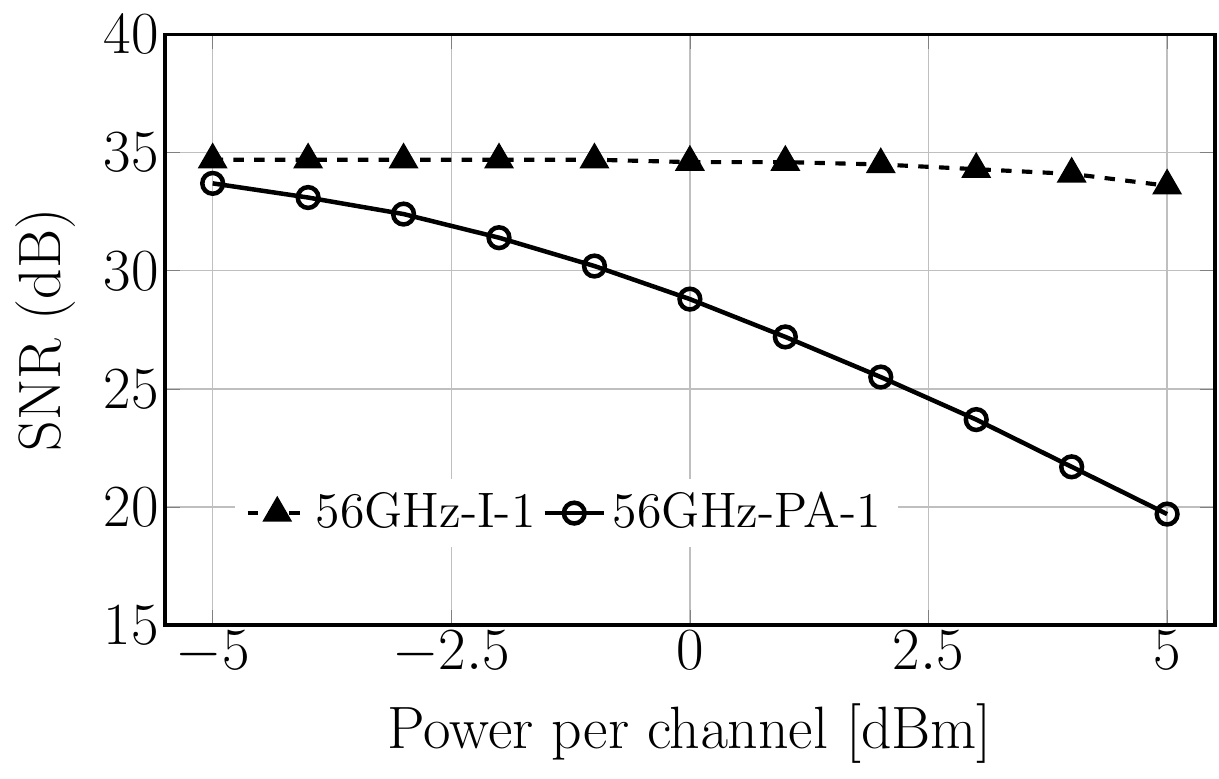}\\
(c)\\
\end{tabular}
    \caption{(a)(b) Simulation diagram of non-integrable models. PA-DBP stands for path-averaged DBP, IDBP for ideal DBP, COI for channel-of-interest. (c) residual distortion of IDBP (shortened in the legend to I, preceded by total bandwidth and followed by number of channels) and PA-DBP (shortened in the legend to PA) systems.}
    \label{fig:DBP}
\end{figure}

In single-channel scenario, if viewed as a compensation scheme, the combination of NFT, back-rotation in non-linear frequency domain, and INFT is somewhat equivalent to the path-averaged DBP. Therefore, we propose to use the residual distortion in a noiseless non-integrable system that is compensated by path-averaged DBP to estimate the approximation distortion. 
Fig.~\ref{fig:DBP}(a)(b) describe two \textit{noiseless} systems with ideal and path-averaged DBP compensation, using the parameters in Tab.~\ref{tab:fibre para}. These systems employ dual polarisation 32-QAM Nyquist signal with 56 GHz bandwidth. In general, any signal whose transceiver DSP causes no distortion suits the purpose. The residual distortion of the system (a) in Fig.~\ref{fig:DBP} is considered equivalent to the approximation distortion in systems of Fig.~\ref{fig:entropy_sym}. Fig.~\ref{fig:DBP}(c) shows that in the power range of our interest ( $\leq-0$ dBm), the approximation distortion is negligible.

\section{Conclusions}\label{sec:conclusion}

The previous work \cite{Yangzhang18ECOC} is extended here with more analysis on the correlation, joint and individual entropy of $b$-modulated signals. The results confirmed the viability and benefits of the $b$-modulation in the DP-NFDM system.

The $q_c$- and $b$-modulated DP-NFDM systems were compared in terms of Q-factor, correlation of sub-carriers, joint and individual entropy.  The $b$-modulated DP-NFDM system shows 1 dB Q-factor improvement over $q_c$-modulated DP-NFDM system due to a weaker correlation of sub-carriers and less effective noise. At last, the $b$-modulated system was optimised for higher data rate, achieving a record net data rate of 400 Gbps (SE of 7.2 bit/s/Hz) on the SSMF with EDFAs over $12\times 80$ km.

\ifCLASSOPTIONcaptionsoff
  \newpage
\fi



%
\bibliographystyle{ieeetr}
\bibliography{sample}

%




\end{document}